\documentstyle[aps]{revtex}

\newcommand {\nc} {\newcommand}
\nc {\ve} [1] {\mbox{\boldmath $#1$}}
\nc {\la} {\mbox{$\langle$}}
\nc {\ra} {\mbox{$\rangle$}}

\begin{document}

\title{Quantum energy flow in atomic ions moving in magnetic fields}
\author{V.S. Melezhik $^{\dag}$ \footnote{permanent address:
Joint Institute for Nuclear Research, Dubna,
Moscow Region 141980, Russian Federation}
and P. Schmelcher $^{\dag\dag}$}
\address{$^{\dag}$ Physique Nucl\'{e}aire Th\'{e}orique et Physique
Math\'{e}matique, C.P. 229, Universit\'{e} Libre de Bruxelles,
B 1050 Brussels, Belgium}
\address{$\dag\dag$ Theoretische Chemie, Physikalisch-Chemisches Institut,
Universit\"at Heidelberg, INF 229, D-69120 Heidelberg,
Federal Republic of Germany}

\date{\today}
\maketitle

\begin{abstract}
Using a combination of semiclassical and recently developed wave
packet propagation techniques we find the quantum self-ionization
process of highly excited ions moving in magnetic fields which has
its origin in the energy transfer from the center of mass to the
electronic motion. It obeys a time scale by orders of magnitude
larger than the corresponding classical process. Importantly a quantum
coherence phenomenon leading to the intermittent behaviour of the
ionization signal is found and analyzed. Universal properties
of the ionization process are established.
\end{abstract}


Rydberg atoms in strong external fields represent an exciting and
very active research area both experimentally as well as theoretically.
During the eighties the main focus was on the hydrogen atom \cite{Fri89} assuming an infinite nuclear
mass which reduces the dimensionality of the system.
However in general the atom possesses a nonvanishing center of mass
(CM) motion in the magnetic field giving rise to a variety of two-body phenomena \cite{Schm97}.
Turning to charged two-body systems like, for example, the $He^+$-ion
the residual coupling of the CM and electronic motion is represented
by an oscillating electric field term yielding five relevant degrees
of freedom. One of the most striking effects caused by the
two-body character of the ion is the recently found classical self-ionization
process \cite{Schm95} which occurs due to energy transfer from the CM to the
electronic motion. Since it is well-known that quantization can severely change the effects
observed in classical dynamics (see ref.\cite{Haa91} and refs. therein)
we develop in the present work a quantum approach to the moving ion in a
magnetic field.

We are interested in the regime of high level density,
i.e. high-lying excitations, for {\it both} the collective as
well as electronic motion which depend on a number of parameters
(field strength, total energy consisiting of the initial CM and
internal energies). The ab initio description of the
quantum dynamics in the above regime goes even beyond modern computational
possibilities and we thus seek a semiclassical approach that is capable of
describing the essential physics of the problem.
The total pseudomomentum \cite{Avr78} is
a conserved quantity associated with the CM motion. In spite of the fact
that its components perpendicular to the magnetic field are not
independent, i.e. do not commute, it can be used to find a suitable
transformation of the Hamiltonian to a particularly simple and physically
appealing form \cite{Schm91,Jo83}. For the $He^{+}$-ion it reads
${\cal H} = {\cal H}_1 + {\cal H}_2 + {\cal H}_3$ where
\begin{equation}
{\cal H}_1 =  \frac{1}{2M}\left({\bf{P}}-\frac{Q}{2}{\bf{B}}\times
{\bf{R}}\right)^2
\end{equation}
\begin{equation}
{\cal H}_2 = \alpha \frac{e}{M}\left({\bf{B}}\times\left({\bf{P}}
-\frac{Q}{2}{\bf{B}}\times {\bf{R}}\right)\right){\bf{r}}
\end{equation}
\begin{equation}
{\cal H}_3 = \frac{1}{2m}\left({\bf{p}}-\frac{e}{2}{\bf{B}}\times
{\bf{r}}+\frac{Q}{2}\frac{m^2}{M^2}{\bf{B}}\times{\bf{r}}\right)^2 \\
+\frac{1}{2M_0} \left({\bf{p}}+\left(\frac{e}{2}-\frac{Q}{2M}\frac
{m}{M}\left(M+M_0\right)\right){\bf{B}}\times
{\bf{r}}\right)^2 -\frac{2e^2}{r} \ .
\end{equation}
Here $m,M_0$ and M are the electron, nuclear and total mass, respectively.
$\alpha=(M_0+2m)/M$ and $Q$ is the net charge of the ion. ${\bf{B}}$ is the
magnetic field vector which is assumed to point along the z-axis.
$({\bf{R}},{\bf{P}})$ and $({\bf{r}},{\bf{p}})$ are the canonical pairs
for the CM and internal motion, respectively. The CM motion parallel
to the magnetic field separates exactly and undergoes a free translational
motion.

${\cal{H}}_1$ and ${\cal{H}}_3$ depend exclusively on the
CM and electronic degrees of freedom, respectively. ${\cal{H}}_1$ describes the
free motion of a CM pseudoparticle with charge Q and mass M.
${\cal{H}}_3$ stands for the electronic motion in the presence of paramagnetic,
diamagnetic as well as Coulomb interactions which, in analogy to the hydrogen
atom \cite{Fri89}, exhibits a variety of classical and quantum
properties with changing parameters, i.e. energy and/or field strength.
${\cal{H}}_2$ contains the coupling between the CM and electronic motion of the
ion and represents a Stark term with a rapidly oscillating electric field
$1/M\left({\bf{B}}\times\left({\bf{P}}- Q/2 {\bf{B}}\times
{\bf{R}}\right)\right)$ determined by the dynamics of the ion.
This coupling term is responsible for the effects and phenomena discussed
in the present investigation.

The essential elements of our semiclassical approach are the
following. Since we consider the case of a rapidly moving ion in a magnetic
field a classical treatment of the CM motion coupled to the quantized
electronic degrees of freedom seems appropriate:
the CM is propagated with an effective Hamiltonian containing the
corresponding expectation values with respect to the electronic quantum states.
The latter obey a time-dependent Schr\"odinger equation which involves
the classical CM trajectory. Both the electronic and CM motion
{\it have to be integrated simultaneously}. The key idea of this semiclassical
approach goes back to refs.\cite{Mcc75,Bil75} where it has been applied
to the dynamics of molecular processes. Our resulting time evolution
equations read therefore as follows
\begin{equation}
\frac{d}{dt} {\bf{P}}(t) = -\frac{\partial}{\partial
{\bf{R}}}{\cal H}_{cl}({\bf{R}}(t),{\bf{P}}(t)) \hspace*{1.0cm}
\frac{d}{d t} {\bf{R}}(t) =
\frac{\partial}{\partial {\bf{P}}}{\cal H}_{cl}({\bf{R}}(t),{\bf{P}}(t))
\hspace*{1.0cm} i\hbar\frac{\partial}{\partial t}
\psi({\bf{r}},t) = {\cal H}_{q}({\bf{R}}(t),{\bf{P}}(t),{\bf{r}})
\psi({\bf{r}},t)
\end{equation}
with the effective Hamiltonian
\begin{equation}
{\cal H}_{cl}({\bf{R}},{\bf{P}}) = {\cal H}_1 + \left\langle \psi({\bf{r}},t)
| {\cal H}_2 + {\cal H}_3 | \psi({\bf{r}},t) \right\rangle \hspace*{1.0cm}
{\cal H}_{q}({\bf{R}}(t),{\bf{P}}(t),{\bf{r}},{\bf{p}}) =
{\cal H}_3 ({\bf{r}},{\bf{p}}) + {\cal H}_2 ({\bf{R}}(t),{\bf{P}}(t),{\bf{r}})
\end{equation}
This scheme represents a balanced treatment of the coupled
classical and quantum degrees of freedom of the ion and takes
account of the energy flow among them.
It possesses the important
property of conserving the total energy which is particularly important
for the correct description of the energy transfer processes occuring
in our system. Since the typical energies associated with the fast heavy CM degrees
of freedom are many orders of magnitude larger than the corresponding elementary
quantum ($\hbar QB/M$) we expect the above scheme to yield reliable results.

Our approach to the solution of the time-dependent Schr\"odinger
equation, which yields the dynamics of an initially defined
wave packet $\psi({\bf{r}},t)$, is based on a recently developed
nonperturbative hybrid method\cite{Mel1,Mel2,MeB}.
It uses a global basis on a subspace grid for the angular variables
($\theta,\phi$) and a variable-step finite-difference approximation
for the radial variable. The angular grid is obtained from the nodes of
the corresponding Gaussian quadrature with respect to $\theta$ and $\phi$,
which is in the spirit of the discrete variable techniques yielding a diagonal
representation for any local interaction\cite{Mel1}.
As a consequence one remains with the Schr\"odinger-type time-dependent radial
equations coupled only through non-diagonal matrix
elements of the kinetic energy operator. This vector equation is propagated
using a splitting-up method\cite{Mar}, which permits a simple diagonalization
procedure for the remaining non-diagonal part\cite{Mel2,MeB}.
Our scheme is unconditionally stable, saves unitarity and has the same order
of accuracy as the conventional Crank-Nickolson algorithm, i.e. $ \sim O(\Delta t^2)$
where $\Delta t$ is the time step size. In order to avoid reflections
of the wave packet from the right edge of the radial grid we introduce
absorbing boundary conditions. The extension of the
radial grid is
chosen (see below) such that it exceeds
the center of the radial distribution of the initial wave packet by more
than one order of magnitude. The typical frequencies associated
with the motion of the Rydberg electron and the CM motion
are different by several orders of magnitude (see ref.\cite{Schm95}).
To investigate the quantum energy transfer mechanisms
requires therefore the integration of the above equations of motion for
a typical time which is a multiple of the time scale of the heavy
particle (CM). This corresponds to many thousand cycles of the
Rydberg electron. Such a detailed investigation would have been
impossible without the use of the above-described combination
of highly efficient techniques.

We assume that the $He^{+}$ ion is accelerated up to some value $E_{CM}$ of
the kinetic energy of the CM motion and its electron is being excited
to some Rydberg state $nlm$ in field-free space. Thereafter it enters the
magnetic field.  In the following we choose $E_{CM}=100 a.u., n=25, l=m=0$ and
a strong laboratory field $B=10^{-4} a.u. (23.5~Tesla)$.
The intial CM velocity is $v_{CM}=0.1656a.u.$ and oriented along the $x-$axis.
We remark that taking the above values for $nlm,B$ for the $He^+$ ion
with the assumption of an infinitely heavy nucleus the corresponding
classical phase space is dominated by chaotic trajectories.
Figs. (1a,1b) illustrate results for the propagation of the wave packet
with increasing time. More precisely we show the intersection
of the integrated quantity $\Psi (\rho,z,t) = \int |\psi|^2 d\phi$
along the cylindrical $\rho$-axis for $z=0$
(fig.1a) and its intersection along the $z$-axis for $\rho=0$ (fig.1b).
Fig. 1a demonstrates that the motion of the wave packet is confined
by the diamagnetic interaction
with respect to the $\rho$-direction, i.e. the direction
perpendicular to the magnetic field. For any propagation time
its value drops by several orders of magnitude at some outer value $\rho_c$
for the $\rho$ coordinate.
As we shall see below (see also figure 3) the variation of  $\rho_c$
is accompanied by a corresponding change in the internal/CM energies
thereby demonstrating the flow of energy between the CM and electronic
degrees of freedom.
Fig. 1b demonstrates that there is for certain time intervals (see below)
almost no decay of the wave packet for large
distances of the $z-$ coordinate. Therefore we encounter a significant flux
of probability parallel to the external field. Reaching the boundary of our
radial grid $r_m=20000 a.u.$ it is absorbed and considered to
represent the ionized state. Having established the existence of an
ionizing probability flux parallel to the magnetic field we immediately
realize from fig.1b that this flux is by no means constant
in time but varies strongly. To see this more explicitly
and also to gain an idea of the overall decay of the wave packet we show
in fig.2 the decay of the norm of the wave packet for a time which roughly
corresponds to one cyclotron period $2\pi M/(QB)$ of the
free CM motion in the field. Fig. 2 shows apart from
an overall monotonous decay of the norm, which is due to the
quantum self-ionization process, an amazing new feature: the
norm exhibits an alternating sequence
of plateaus and phases of strong decay. The widths of the plateaus slightly
increases with increasing time. This intermittent behaviour of the
ionization signal from the moving ions is a pure quantum phenomenon,
i.e. does not occur in the corresponding classical ionization rates
\cite{Schm95}. Furthermore the calculation of the classical ionization rates
for the same parameters (field strength, energies) yields a typical ionization time
which is by two orders of magnitude smaller than the one obtained by the
quantum calculation. The ionization process
is therefore significantly slowed down through the quantization of the
system which is in the spirit of the quantum localization processes
shown to exist in a variety of different physical systems (see ref.\cite{Haa91}
and refs. therein).
The observed slowing down of the ionization process represents one important
difference of the classical and quantum behaviour of the moving ion which
occurs in spite of the fact that we are dealing with a highly excited system.

The obvious question of the origin of the intermittent occurrence
of the plateaus and ionization bursts arises now. At this
point it is helpful to consider the behaviour of the CM energy as a
function of time which is illustrated in fig. 3.
Starting with $E_{CM}=100a.u.$ at $t=0$ we observe
a fast drop of it for short times yielding a minimum of $E_{CM}$
at approximately $t_1=4 \times 10^{7}a.u.$. Thereafter it raises and reaches
a maximum at approximately $t_2=1.2 \times 10^{8}a.u.$ after which it drops
again, i.e. it shows an overall oscillating behaviour. The ``valleys'' of
$E_{CM}$ coincide with the plateaus of the norm decay whereas the regions with higher
CM velocities correspond to phases of a strong norm
decay of the wave packet. The increase of the widths of the plateaus
in the norm decay (see fig.2) matches the
corresponding decrease of the frequency of the oscillations of the CM energy.
Since the total energy is conserved this clearly shows that the
ionization bursts correspond to phases of relatively low internal energy
(although certainly above the ionization threshold) whereas the phases of
higher internal energy go along with the plateaus of the norm behaviour,
i.e. the localization of the electronic motion. This provides the key for
the understanding of the rich structure of the norm decay.
The phase of high energy for the electronic motion means that the magnetic interaction
strongly dominates over the Coulomb interaction. This makes the electronic motion
approximately separable with respect to the motion perpendicular and parallel
to the magnetic field. As a consequence the energy transfer process from
the degrees of freedom perpendicular to those parallel to the external
field are very weak i.e. the ionization process is strongly suppressed.
This corresponds to an almost integrable situation
for the ions dynamics. On the contrary for relatively
low internal energies the Coulomb interaction is much more relevant and
mediates together with the coupling Hamiltonian ${\cal H}_2$ the energy
transfer from the CM to the electron motion parallel to the magnetic
field. As a result we encounter a flow of probability in the +/-z-directions
which corresponds to the ionization burst.
During this period
of motion a comparatively strong dephasing of the wave packet takes place.

The intermittent behaviour of the ionization rate can therefore be
seen as a quantum manifestation for the switching between different regimes
of the internal energy corresponding to weaker or stronger Coulomb interaction.
Pumping energy from the CM to the electronic motion weakens the Coulomb
interaction and leads to the suppression of the ionization process whereas pushing the
energy back to the CM motion decreases the internal energy and enhances
the Coulomb interaction. To elucidate the time scale on which this process takes
place we have computed the autocorrelation function $C(t)=<\psi(t)|\psi(0)>$
where $<>$ means integration over the electronic coordinates. As a result
we observe a modulation and recurrence of the autocorrelation at a
time scale $t\approx 1.6 \times 10^{8}a.u.$ which corresponds approximately to the
recurrence of the plateaus for the norm decay. The corresponding power spectrum
shows a broad peak at a frequency $\omega \approx 3.5\times 10^{-8}a.u.$.
An important feature of the quantum self-ionization
process is the approximate stability of the time intervals corresponding
to the plateaus of the norm (no ionization signal) with respect to variations
of the initial CM velocity of the ion. Our investigation shows that decreasing the CM energy from
100$a.u.$ to 12$a.u.$ leads to a decrease with respect to the distances between the plateaus,
i.e. the difference of the norm values belonging to different plateaus,
roughly by a factor of two. This corresponds to a
significant slowing down of the ionization process. However the widths of the plateaus
remain rather stable and {\it represent therefore a universal quantity}
which is approximately independent of the CM velocity.
Varying the field strength causes a change of both
the distances between the plateaus and their widths.

The quantum self-ionization process should have implications
on the physics of atoms and plasmas occurring in a number of different circumstances.
Apart from this it obviously suggests itself for a laboratory experiment
(the lifetime of the Rydberg states exceeds the
time scale of ionization by orders of magnitude)
which should be very attractive due to the expected
intermittent ionization signal which is a process revealing the
intrinsic structure and dynamics of the system during its different
phases of motion.

This work was supported by the National Science Foundation
through a grant (P.S.) for the Institute for Theoretical
Atomic and Molecular Physics at Harvard University and
Smithsonian Astrophysical Observatory. P.S. thanks H.D.Meyer and D.Leitner
for fruitful discussions.
V.S.M. gratefully acknowledges the use of the computer resources of the IMEP of
the Austrian Academy of Sciences, he also thanks the PNTPM group of
the Universit\'e Libre de Bruxelles for warm hospitality and support.

\vspace*{1.0cm}

\begin{center}
{\bf Figure Captions}
\end{center}

{\bf Figure 1:} (a) The intersection $\Psi (\rho,z=0,t) = \int |\psi|^2 d\phi$
along the $\rho$ axis. (b) The intersection  $\Psi (\rho=0,z,t)$ along the $z$ axis.
(Atomic units are used).

\vspace{0.5cm}

{\bf Figure 2:} The norm of the electronic wave packet as a function of time
(in units of $10^{8}$ atomic units).

\vspace{0.5cm}

{\bf Figure 3:} The CM energy as a function of time
(in units of $10^{8}$ atomic units).


\vspace*{2.0cm}

\begin{thebibliography}{}

\bibitem{Fri89} H.Friedrich and D.Wintgen, Phys.Rep. {\bf{183}}, 37 (1989)
\bibitem{Schm97} P.Schmelcher and L.S.Cederbaum, 'Atoms and Molecules in
Intense Fields', {\bf Springer Series}: Structure and Bonding {\bf{86}},
27 1997
\bibitem{Schm95} P.Schmelcher and L.S.Cederbaum, Phys.Rev.Lett.{\bf 74},
662 (1995); P.Schmelcher, Phys.Rev.A {\bf 52}, 130 (1995)
\bibitem{Haa91} F.Haake, 'Quantum Signatures of Chaos', Springer Verlag 1991.
\bibitem{Avr78} J.E.Avron, I.W.Herbst and B.Simon, Ann.Phys.(NY) {\bf{114}},
431 (1978)
\bibitem{Schm91} P.Schmelcher and L.S.Cederbaum, Phys.Rev.A {\bf 43}, 287 (1991)
\bibitem{Jo83} B.R.Johnson, J.O.Hirschfelder and K.H.Yang, Rev.Mod.Phys.
{\bf{55}}, 109 (1983)
\bibitem{Mcc75} K.J.McCann and M.R.Flannery, Chem.Phys.Lett.{\bf 35},
124 (1975); J.Chem.Phys.{\bf 63}, 4695 (1975)
\bibitem{Bil75} G.D.Billing, Chem.Phys.{\bf 9}, 359 (1975)
\bibitem{Mel1} V.S.Melezhik, Phys.Rev.A {\bf 48}, 4528 (1993);
\bibitem{Mel2} V.S.Melezhik, 'Atoms and Molecules in Strong External
Fields', p.89, edited by P.Schmelcher and W.Schweizer, Plenum Publishing
Company 1998.
\bibitem{MeB} V.S.Melezhik and D.Baye, Phys.Rev.C (in press)
\bibitem{Mar}
G.I.Marchuk, On the theory of the splitting-up method, in
{\it Partial Differential Equations. II. SYNSPADE-1970} (Academic,
New York, 1971)

\end{thebibliography}
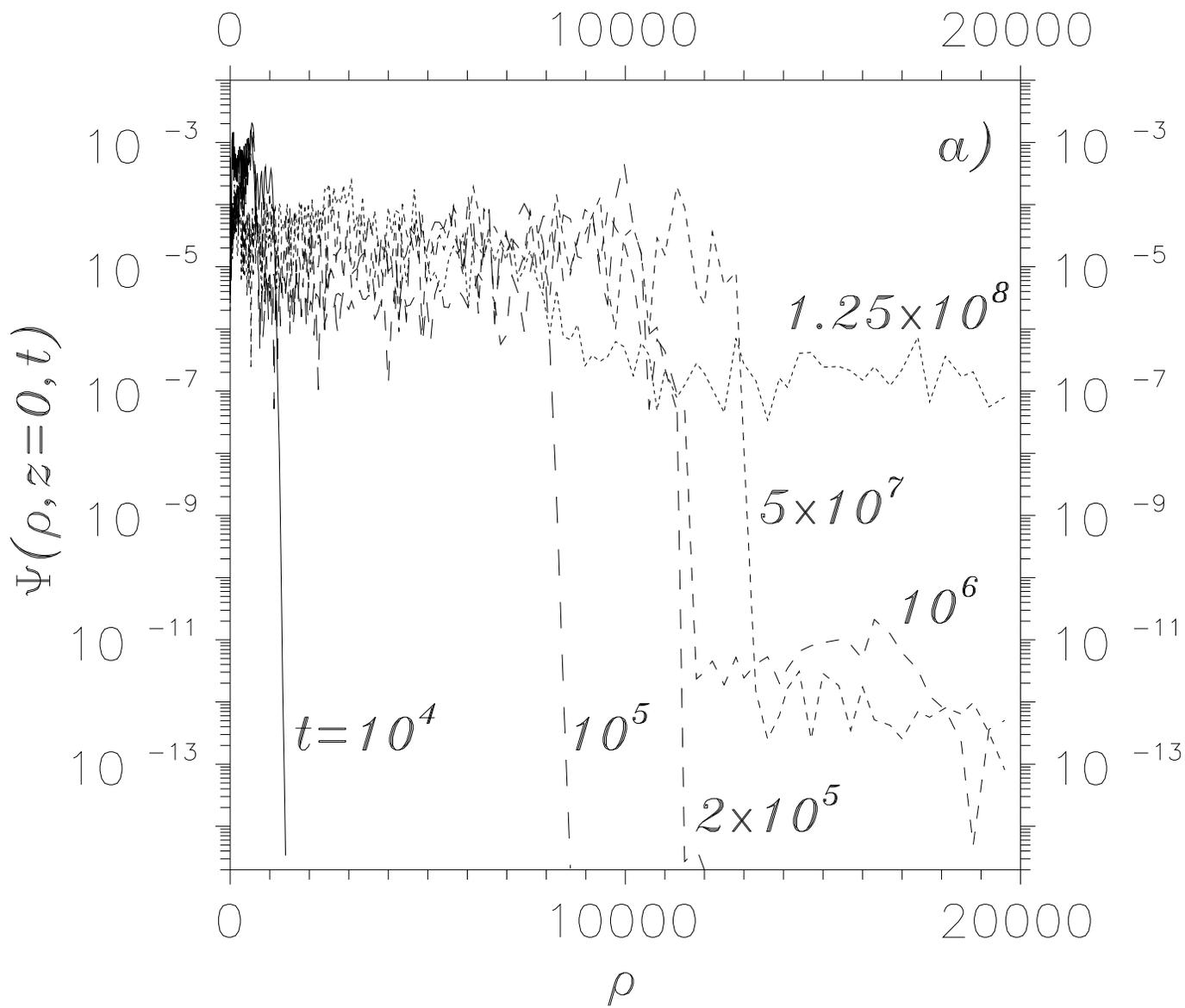
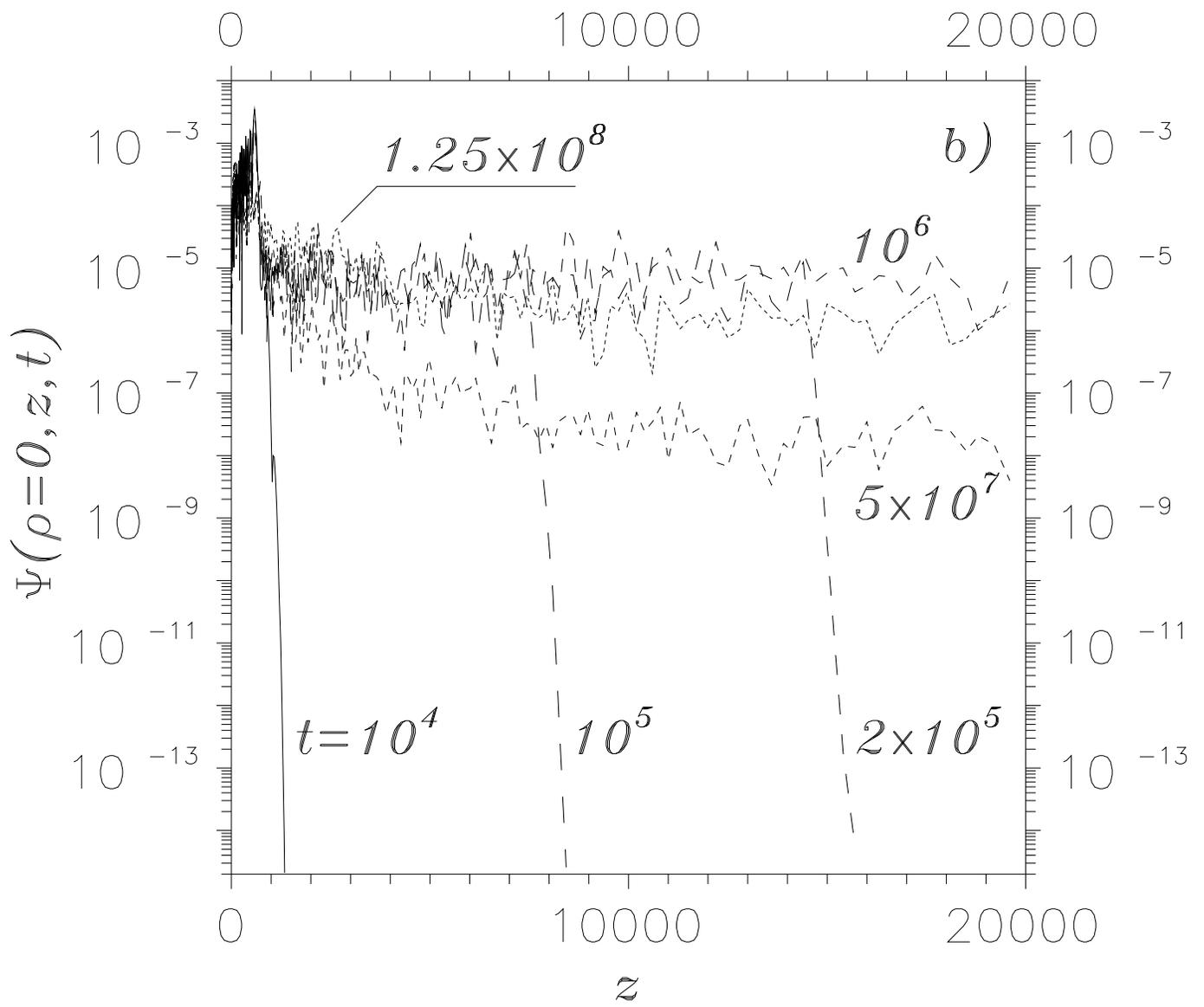
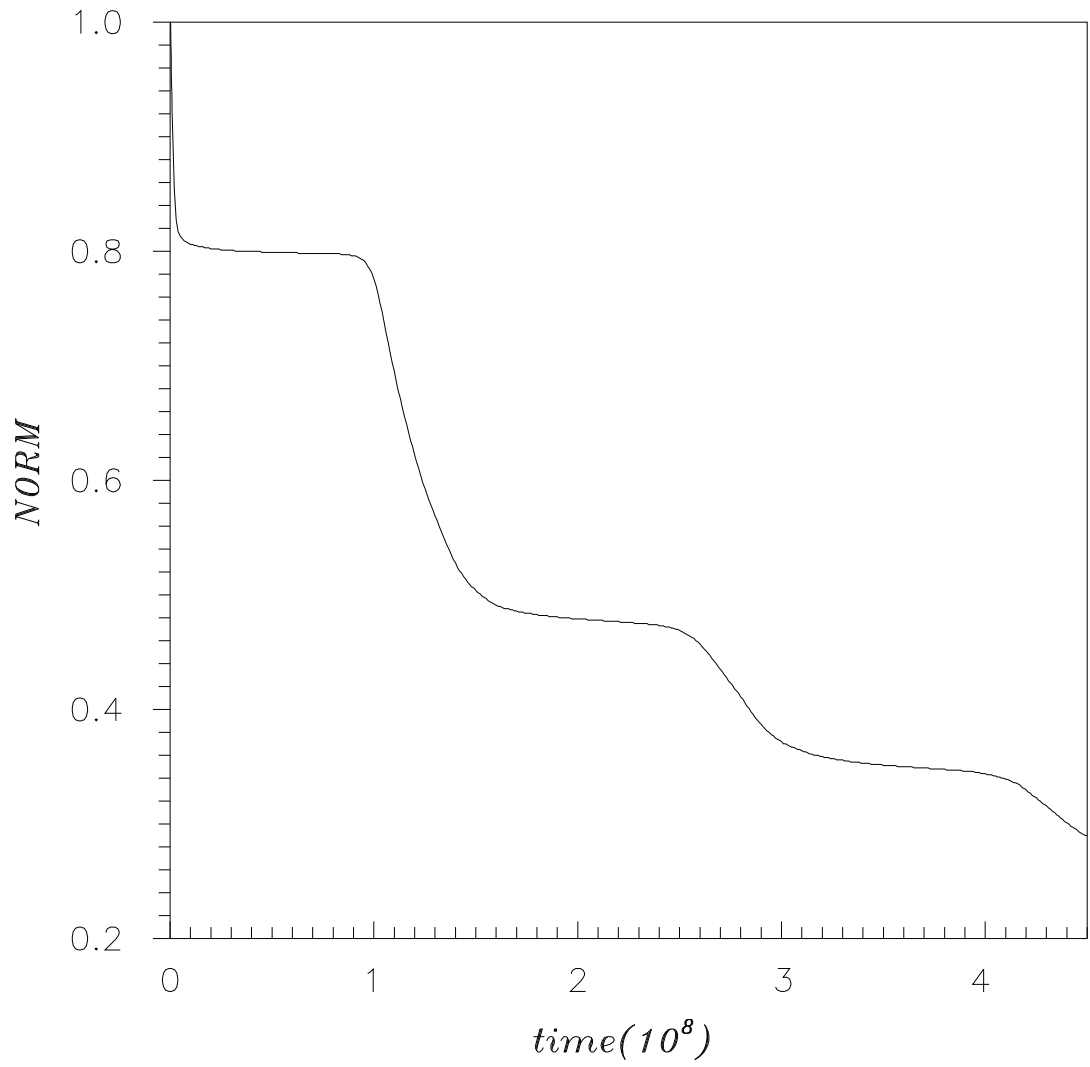
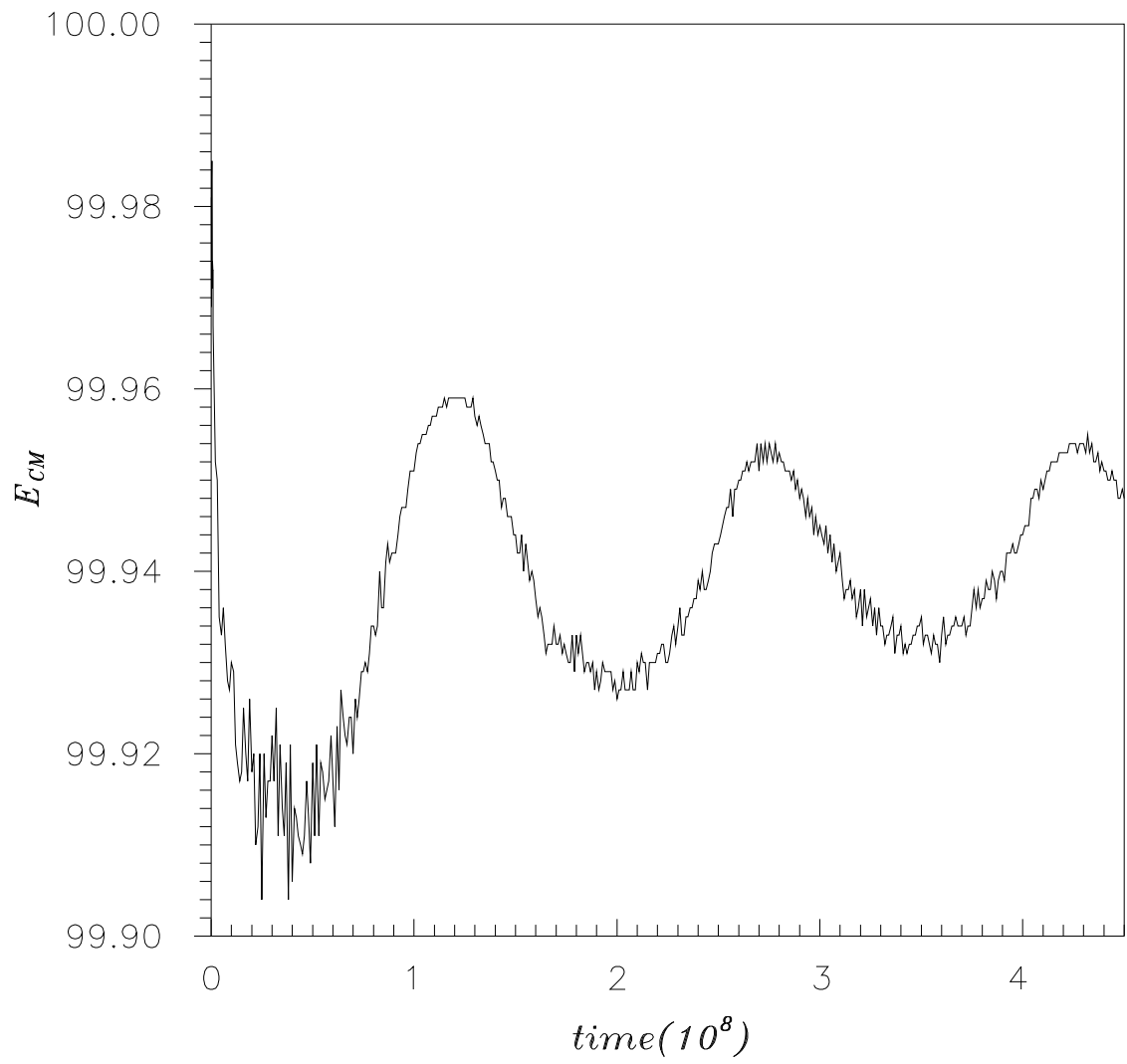
\end{document}